# 5PEN TECHNOLOGY: A NEW DAWN IN HOMOGENEOUS AND HETEROGENEOUS COMPUTING

Osagie Scale Uwadia Maxwell [1], K.O. Obahiagbon[1], Osagie Joy Amenze[2] and John-Otumu M. A.[2]

[1] Department of Mathematics and Computer Science, Faculty of Basic and Applied Sciences Benson Idahosa University, G.R.A., Benin City, Edo State, Nigeria
[2] M.Sc Student: Department of Computer Science, School of Post Graduate Studies, Benson Idahosa University, G.R.A., Edo State, Benin City, Nigeria
[2] Department of Computer Science, Ambrose Alli University, Edo State, Ekpoma, Nigeria

## *ABSTRACT*

*This research work is a pair review into the conceptual frame work and innovation into Pen-style Personal Network Gadget Package (P-ISM) as inevitable tool to easy, fast and convenient access to the internet. Computing activities have increased the degree of people using personal computers (PCs), complicated packages and all form of social media applications (Apps.) have emerged within this short period. Meeting these trends (day to day activities) in more convenient form has led to the modern sophisticated garget such as Pen-Style Network Gadget Package (P-ISM) prototype. The growth in internet affects our lives in much better way than we know and its sustainability made 5 pen technology innovations a salt after.*

## 1. INTRODUCTION

Late 18th and early 19th centuries were the era of divided information and this was as a result of unsophisticated machines and enormous heat generated by the machines, the non availability of network made this period stringent and complex. Various countries at this period operated separately and information became scarce due to systems engender, those with useful information made an empire around it and became emperor, thereby making those without subjects. Though still in existence, not commensurate to 18th and 19th century. This period also revealed that technology advancements can attain full height if human become pragmatic and empirical about her society. [1]

With the exponential increase in internet users, it can be attested to that major progresses have been made by scientists and engineers in breaking the communication barriers via modern day technologies. Ironically, the more difficult the activities of humans are, the more challenged are scientists and engineers in coming up with new and sophisticated technology capable of addressing the challenges. The rise in availability of internet has mad the world a global village and has enable sharing of knowledge by scientist on the best possible way to resolving and supporting fellow partners in getting viable and possible solutions to the challenges faced by humans. [2]

International Journal of Computer Science & Information Technology (IJCSIT) Vol 8, No 2, April 2016

Virtually everything done by human has two sides (Positive and Negative). But, on the quest to rendering services to minimize the negative is one of the greatest challenges faced in today's changing world. The question is why 5pen technology?

The technological foundation of computer is to help bridge the gap between human and labour. Following human's inability in maintaining setting positions for a period of time has led to scientists and engineers coming up with an idea of a more portable system called 5pens technology. This technology addresses human's inability in maintaining setting positions.

### 1.2. Research Aim and Objectives

The overall aim of this research work is to carry out study on 5Pen Technology: A New Dawn in Homogeneous and Heterogeneous Computing. The objectives are as follows:

1) To review the notion (concept) frame work, and the components of this technology
2) To educated the public the need to further embrace this technology
3) To bring to the understanding the need for collaboration between researchers, IT professionals, Scientists and Engineers in bringing about the realisation of this technology
4) To make know the safety and convenient nature of this technology

## 2. LITERATURE

### 2.1. Advent of 5pen Technology

5Pen Technology Just like any other technology has its history dated back to the introduction of digital computer where computer started utilizing the coherent principle of structural (segmented) view function and a defined attribute. The structure of digital computer involves Input phase, Output phase, Processing phase, Memory phase, Control unit etc. Since the introduction of digital computer as described by Von Neumann digital computer systems, the designed in this format/structure has experienced a turnaround progress unlike the analogue days where systems were error prone. Retrieving and transmitting data in this era were known to be difficult and the flexibility of the systems designed was in doubt. This period actually made resources sharing more difficult leading to the slavering of people without information.

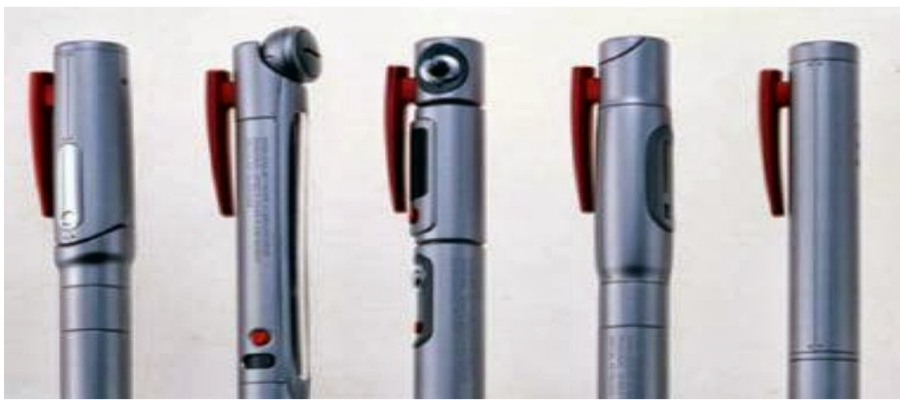

Figure 1. Picture of 5Pen Technology.

The actual introduction of 5pen technology was at the 2003 ITU Telecom World exhibition that took place in Geneva, where NEC Corporation made known a tangible prototype named "Pen-style Personal Networking Gadget Package," or P-ISM, and is currently gaining more support in the world of information and communication technology (ICT). [3], [4], [5]

NEC Corporation is a multinational company who specialises in the production of information technology equipments. NEC has more than two headquarters in different countries and they are still expanding their tent in IT world. Majority of the equipments use in communicating via a network are tied to the history of NEC foundation, a cooperation who over the years has been the salt after in information communication technology related matter. NEC cooperation has over the years maintained integrity in IT industries. Today some other mini IT industries have learnt to use NEC Corporation as model in achieving their desire design. Today various gargets enabling information communications (IC) have evolved. People today view or make contributions to the advancement of technology as required by various countries because computing has become ubiquitous. [6]

## 2.2. The Goal

The world of computing over the past three decades has experienced sweet shift, this has actually surpassed human projection. The computing world is still experiencing revolution. The aim of 5pen technology is to minimize/eradicate human most difficult problem, that is, human inability in maintaining stable position within a period of time and the inconveniences in travelling, moving and accessing all-round information with gadgets such as laptops, iPads, tablets etc.

The lay down approach by NEC Corporation in the introduction of prototype called 5Pen Technology at Geneva ITU telecom exhibition can be seen as a major breakthrough in the mobile industries and this technology should be perpetually given full support by other IT industries in bringing about the full realisation [6]

A critical view into what has become heterogeneous and homogeneous computing today started in the pre-industrial age. As information was transmitted over line-of-sight distance (The letter, thereafter extended with telescope) using signal smoke, torch signally, flashing mirrors, signal flares/semaphore flags. This technology took a new shape with the introduction of wireless. The little wireless concept today has revolutionised the telecommunication industries. Samuel Morse in 1838 developed set of signals that was later replaced in 1895 by Guglielmo Marconi telephone concept. The idea was to covey complex messages, but was thereafter change to the radio packet network (ALOHANET) which gave impetus to the wide area wireless data service in 1990 and was developed in the University of Hawaii in 1971. Over the years this technology has helped to bridge the divided world of information around the globe with return on investment.[7], [8].

5 pen technologies has the same background with the emergence of wireless service in the pre-industrial era which in question provides convenience. The technological tool should not be underestimated by professional bodies and all well grounded IT firm should as a matter of expediency create a research body who can see to the full implementation rather than leave the idea and the concept with NEC cooperation. All concerns individual should expedite actions it realisation so as to eradicate the difficulties face in moving laptops, tabs, ipads, desktops etc along.[9]

## 2.3. 5Pen Classifications

The flexibility of 5Pen Technology as demonstrated in Geneva Telecom Exhibition by NEC Corporation in 2003 is the ability of the technology in combining already existing technologies such as Bluetooth, Network, Camera and virtual reality techniques into the full realization of the prototype. The diagram in figure 2.0 shows each class of the 5Pen with their distinct function.

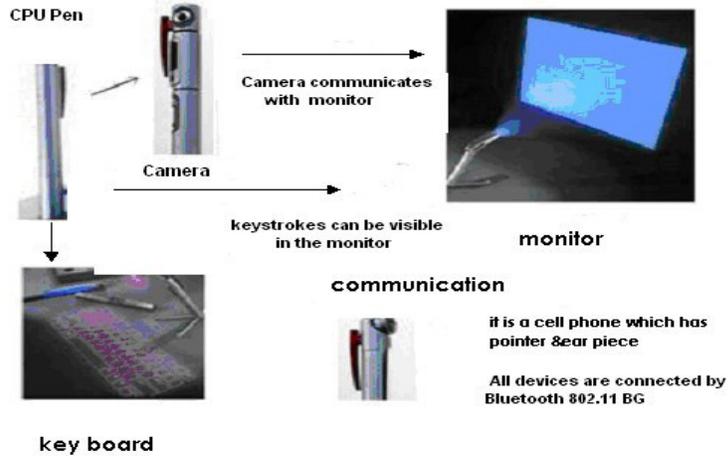

Figure: 2. Classification of 5Pen Technology

The heterogeneous and homogenous operations of 5pen are inter-dependent. Each performs a function required by the others. 5Pens combines series of technologies in bringing about complete computer system that has the ability/capability of doing same work as office computer (Desktop/Laptop) but in this case, there is no restriction of movement. A typical example to the functionality of this gadget is the human body where each part of the body provides special service to the easy movement of the body. This technology does not grant a Pen supremacy (one of the pen) over others (Pens) and the 5Pens being dependent entities; rely solely on each other to form a complete computer [10], [11]

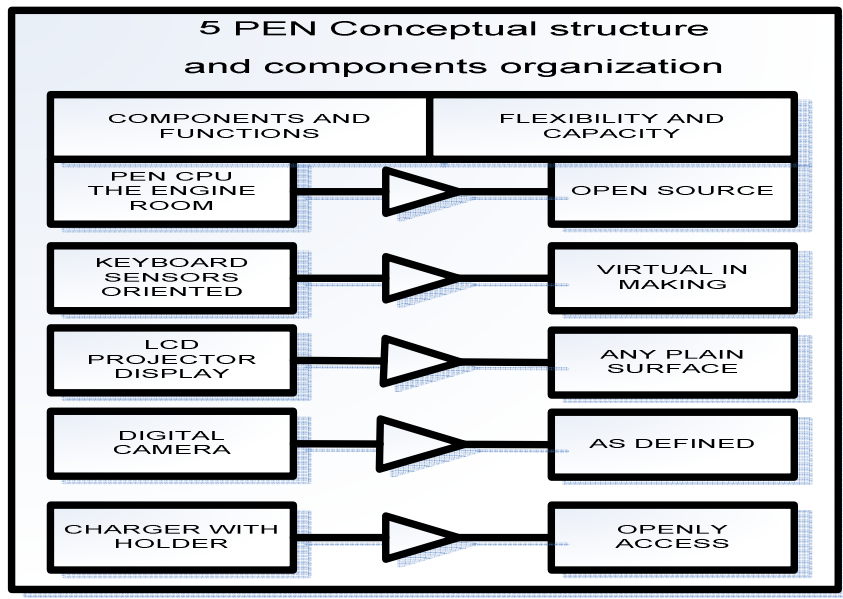

Figure 3. Functions of each component of 5 Pen Technology and it conceptual frame work

International Journal of Computer Science & Information Technology (IJCSIT) Vol 8, No 2, April 2016

**2.3.1 Engine Room (CPU)**

The engine room which is also the Central Processing Unit (CPU) is the part of the computer systems that carries out all processes, along with taking care of input and output data for onward computation and evaluation. One of the 5Pens is responsible for this part and does all functions of computing relating to logical process. Every instruction and program of the computer is coordinated by this pen. This session of computer is involved in the logical operations, input, output and arithmetic computation and its movements is strictly on electrical signal, and this gives command to computer for the execution of programs and instructions. Long before the introduction of microprocessor, many computers worked on Integrated circuit (IC). Microprocessors are the integration of many IC which brought about the implicit speed in processing rate. This is as a result of the elimination in energy consumed and the reduction in heat generation.

This approach has helped in the designer of Megabytes and Gigabytes of microprocessors that is revolutionizing the information communication technology industries. The fetching, decoding, executing, and writing back is the operational approach of CPU in instruction execution. Every instruction is tied to the CPU clock rate and this determined the speed of execution of a given instructions. [12]

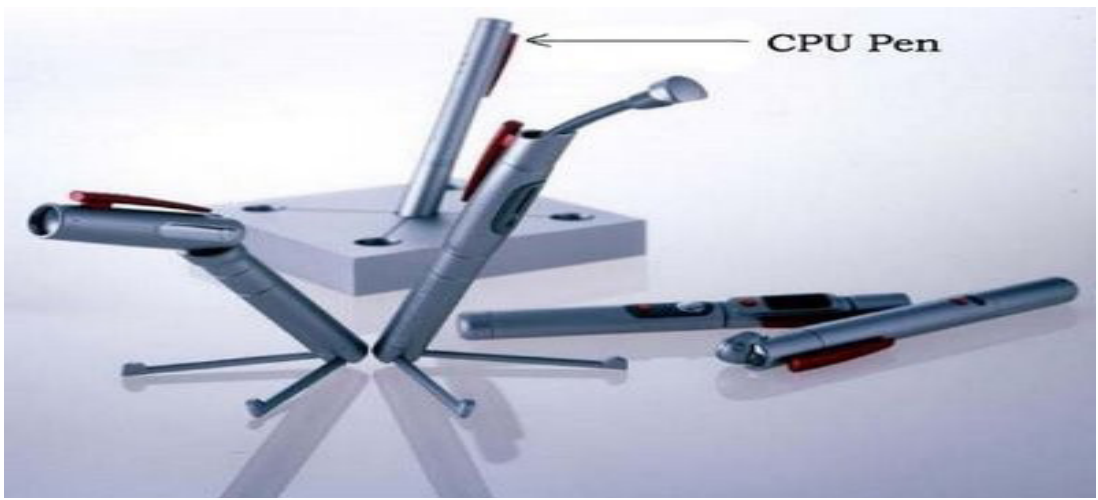

Figure: 4. Diagram of Pen PC CPU

**2.3.2. Virtual Keyboard**

Part of the world sees VK as old technology while others such as Africa considered this technology new and innovative in computing world. Virtual keyboard is a technology that has come to replace the physical keyboard or the unscreened keyboard. The technology looks so simple but has embedded complexity in it combination. The operational function includes laser discharge. The beam released is focused (projected) on a plain table thus, looking as if the virtual keyboard is denoting a standard layout keys for typing in which the leftmost keys of the top lettered row are Q.W.E.R.T.Y. The information needed by the computer is communicated to the centre processing unit through the projected keyboard (virtual keyboard). The beam of light is usually projected to a flat surface where adequate typing can be done and the connection to the main component is through a short range Bluetooth.

Virtual keyboard in this technology is inbuilt software that enables input data. Unlike the hardware keyboard, virtual keyboard does not require cable. The connection is via wireless channel (Bluetooth). Methods to which this input are done varies from one level to another and this includes all current manner at which data are inputted into the system (screen touch, speech etc.) [12]

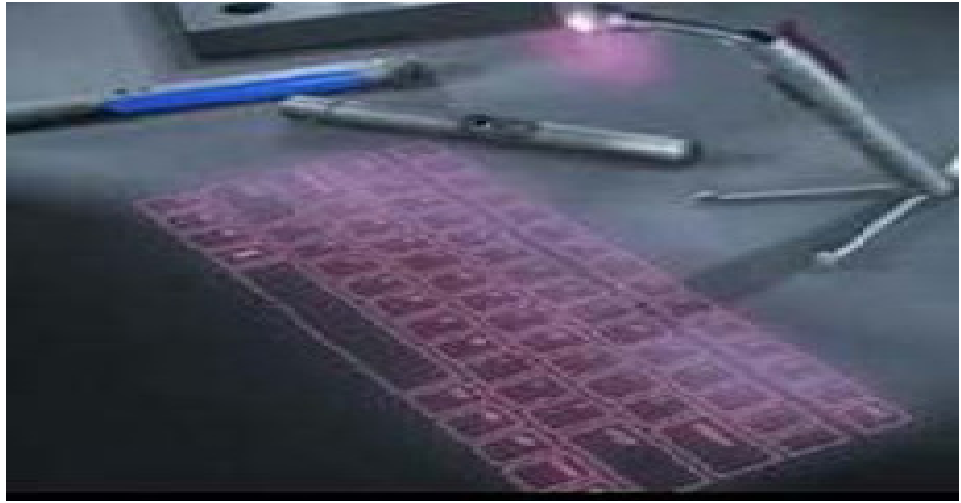

Figure 5. Virtual Keyboard

### 2.3.3. Light Emitting Diode (Led) Projector

The light emitting diode (LED) projector creates a projection screen to which the work or computing carried out via the virtual keyboard can be viewed. It has a well suitable resolution of 1024 x 768. The projector functionality is directly proportional to the images display via the input device and has clearer resolution of all images projected through it. The projector allows for document/research presentation, delivering of lectures, seminar presentations etc.

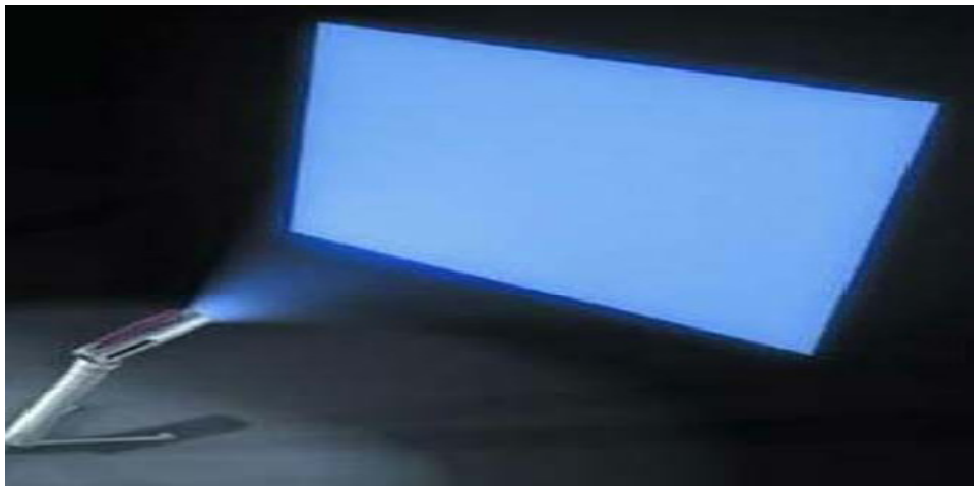

Figure: 5. Light Emitting Diode (LED) Projectors.

**2.3.4. Digital Camera**

One of the 5pens is use as camera in capturing images that are stored in the internal memory. Most digital camera synchronises the environment so as to replicate the image as real as possible. The image captured by this pen is sent to the centre processing unit (CPU) via an intermediate network channel. Today various researches have further enhanced the quality of digital cameras seen today. The more study carried out into augmented reality will bring about quality and coherent images that truly represent the true nature of such images. The camera is considered important in the areas of video recording, conferencing; online chatting etc. The camera is fully regarded as web cam, and uses Bluetooth to interact or connect other devices. The camera works on a full circle (360 degree) shape in capturing it surroundings.

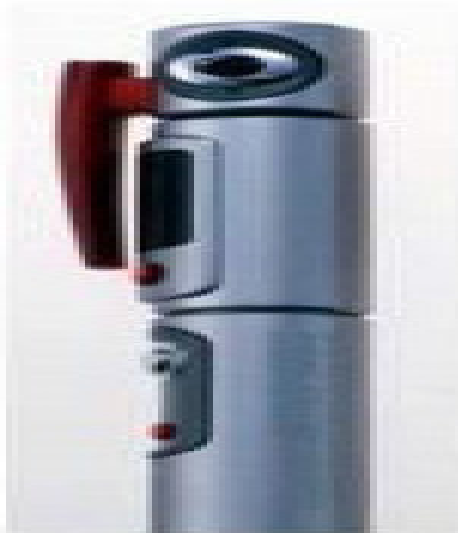

Figure 6.5Pen technology Digital Camera

**2.4. Other Considerable Factors**

Any information communication technology developed has one major aim and this is to help share or disseminate information. Disseminating information cut across using the technology along network paths and this could be international (WAN) or Local (LAN). No matter how sophisticated the technology could be it must adhere to the model (OSI) of modern day communication so as to allow easy flow or exchange of important data or resources. Open System Interconnection (OSI) is a model that demonstrates how all communication gadgets communicate freely irrespective of the type, brand, and architecture, any system without adherence is by default engender. *See OSI reference model for detail* [13]

Ravinder S. Kang 2001 in his work titled *Data Communications Networking* defines networks

> ➢ *" A collection of entities wanting to share and have a conversation"*
> ➢ *"The exchange of messages,information or data through physical sources"*
> ➢ *"Resources or facilities that make that conversation"* [14],

The under laying meaning of this definition stood on the fundamental principle of the philosophical stands of hardware and software standard in data exchange. The physical entity in the above definitions represents all form of hardware requirements; the entity collection is brought about by a synchronised software application through the agreed protocol that does the handshaking and signalling [15]

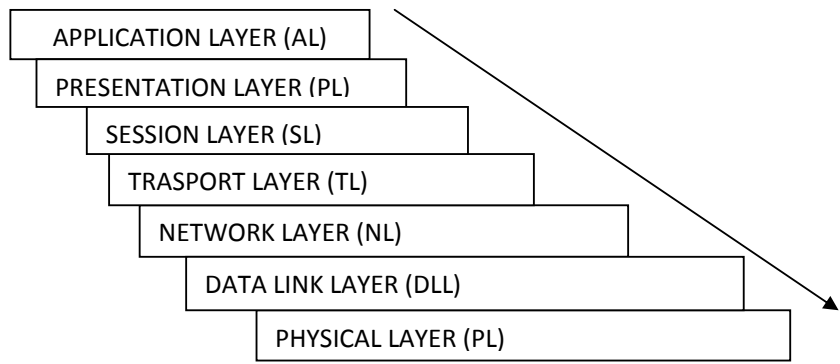

Figure 7. Open System Interconnection (OSI) MODEL

### 2.4.1. Bluetooth

Bluetooth is a mechanism used in exchanging data within a given range says 2400 to 2484 MHz. the bands (spectrum) create a structural platform (radio frequency) for which all Bluetooth devices functioned. Activated Bluetooth enable exchange of signals or data within accessible limit. In some operating systems such as Android access time is considered to be factor. The range specified above is free licensed spectrum by Industrial Scientific and Medical (ISM). The Bluetooth and the entire gadgets are connected to the internet via cellular phone function.

### 2.4.2. IEEE 802.11

To every communication there exist rules and guidelines to establishing communication. Open System Interconnection (OSI) model laid more emphases on how information/data transmitted via a channel travel from sending machine (A) to receiving gadget (B). The layers in figure: 7 shows how an X signal voltage is set up on the sending gadget (A) and acknowledgement received that information delivered to receiving gadget (B) that the X signal voltage has just be set up. There are other well known bodies that have helped to enhance the computing standard by way of enacting standard laws to govern the way and manner in which devices communicate. One of these bodies is the Institute of Electrical Electronic Engineering (IEEE). The standard for allowing communication establishment in gadget is the IEEE 802. 11. The frequency for establishment is from 2-5GHz band. [16], [17].

### 2.4.3. Cellular Network

The effectiveness of handheld devices is on its ability in switch network. 5pen technology communicates with other handheld devices via integrated cellular systems. A cellular system or network is a radio network with a base station fixed at different location covering large geographical area. Each base station has the capacity of further enhancing the quality of the network received and retransmits it from one region to another (Topology Dependent). The base station handheld allows handheld devices that are remotely located to have access to the network. e.g. Laptops, Mobile phones, 5pen technology etc

**2.4.4. The Battery**

The most considerable part of portable devices (mobile devices) such as 5pen is the battery and the storage strength. The longer the battery the more people appreciate it usage. Batteries are meant to work longer for a working hours but this largely depends on the usage of the gadget. For 5pen, the battery is lithium ion battery and has the capacity to minimize energy. It is cost effective. Durability remains one of the greatest strength of lithium ion batteries. [18]

## 3. AUTHENTICATION

### 3.1. Security

The greatest concern by all end users in any newly introduced gadget is the security features i.e. how well the system or machine designed would protect confidentiality. [19], [20], [21], [22]. 5pen has a unique security features that made it one of the most secured portable devices. The e-fingerprint mechanism matches user biometric data with the database before granting authorisation. The technology is not just portable, is efficient, and user friendly. It doesn't require tedious training before using it and it was said in the introductory part of this research that a prototype has been developed in 2003 and is expected that more collaborations by information technology professionals will bring about the realisation of this technology in coming years. For details read. [23], [24], [25]

### 3.2. Advantages of 5pen Technology

The advantages of this technology are endless. The portable nature of the technology is convenient, feasible and ubiquitous (heterogeneous). The technology was developed with cellular phone technology which makes it internet friendly, the mobile and the virtual architecture is another huge and important aspect of 5pen technology. Aside the above listed advantages, the technological ability of 5 pen technology cubing internet crime through the biometric authentication remains one of the greatest strength of P-ISM [26], [27].

## 4. CONCLUSION

The increase amount in sophisticated gargets currently in the market and the rate at which people patronized these gargets gives clear indications of PC impacts amongst end users. The compact nature and the signal strength of this innovation (P-ISM) as demonstrated At the ITU Telecom World exhibition held in Geneva 2003 by Tokyo-based NEC Corporation of a conceptual prototype of what they dubbed a "Pen-style Personal Networking Gadget Package," (P-ISM) unveil a new era in information communication technology advancements. The own idea is geared towards full realisation of easy, fast and convenient heterogeneous and homogeneous computing. Our view from the above concept is the enrichment of political, economical and selfless willpower in its (P-ISM) realization.

As more developments in IT continues, critical examinations must be made on the flexibility of 5 pen technology prototype, so as to create room for the reengineering of the key elements (P-ISM) into full usage. It is believed from this research that the reengineering of this prototype (P-ISM) into full realization will change the scope of information communication technology and will give return on investment.

P-ISM is a gadget package including five distinct functions as it was articulated above. A pen-style CPU, Projector, Cellular phone function, Virtual keyboard, and Camera, connected together

International Journal of Computer Science & Information Technology (IJCSIT) Vol 8, No 2, April 2016

with short-range wireless technology (Bluetooth) and via a cellular phone (internet).  The entire pen recharges its batteries with holding block and holds the mass storage. Each pen communicates wirelessly,

This research has show via a review that hitherto, humans have had issues with moving around with current information communication technology gadgets, thereby leading to frustrations. Our result is so clear and straight forward.  Knowing that the world is constantly going trending, it will be wise and profitable if all professionals, technocrat, engineers and scientist form a collaborated idealistic and realistic force in ensuring its (P-ISM) implementation. This will eradicate current challenges associated with the mode and manner at with computing is been done and this was fully captured by the goal of this research above.